\documentclass[conference]{IEEEtran}
\IEEEoverridecommandlockouts

\usepackage{cite}
\usepackage{amsmath,amssymb,amsfonts}
\usepackage{hyperref}
\hypersetup{colorlinks=true, unicode=true, linkcolor=[rgb]{0.10,0.05,0.67}, citecolor=[rgb]{0.10,0.05,0.67}, filecolor=[rgb]{0.10,0.05,0.67}, urlcolor=[rgb]{0.10,0.05,0.67}}
\usepackage{algorithm}
\usepackage{algorithmicx}
\usepackage{algcompatible}
\usepackage{algpseudocode}
\usepackage{amsmath,array,graphicx}
\usepackage{soul,xcolor}
\usepackage{float}
\usepackage{tikz}
\usetikzlibrary{decorations.pathreplacing,calc}
\newcommand{\tikzmark}[1]{\tikz[overlay,remember picture] \node (#1) {};}
\setstcolor{red}

\usepackage{subfigure}
\usepackage{tikz}
\usetikzlibrary{backgrounds}
\usepackage{xcolor}
\usepackage{pgfplots, amsmath}

\algdef{SE}[DOWHILE]{Do}{doWhile}{\algorithmicdo}[1]{\algorithmicwhile\ #1}%
\algnewcommand\algorithmicforeach{\textbf{for each}}
\algdef{S}[FOR]{ForEach}[1]{\algorithmicforeach\ #1\ \algorithmicdo}

\usepackage{dblfloatfix}
\usepackage{hhline}
\usepackage{array, makecell, cellspace}
\usepackage{graphicx}
\usepackage{textcomp}
\usepackage{xcolor}
\usepackage{multirow}
\def\BibTeX{{\rm B\kern-.05em{\sc i\kern-.025em b}\kern-.08em
    T\kern-.1667em\lower.7ex\hbox{E}\kern-.125emX}}
\pgfplotsset{compat=1.18} 
\begin{document}

\newcommand\Aysucomments[1]{\textcolor{red}{#1}}
\newcommand\Aysucomment[1]{\textcolor{red}{#1}}
\newcommand\Emrecomments[1]{\textcolor{blue}{#1}}
\newcommand\Emrecomment[1]{\textcolor{blue}{#1}}
\newcommand\Arsalancomment[1]{\textcolor{brown}{#1}}
\newcommand\Arsalancomments[1]{\textcolor{brown}{#1}}

\newcommand\Seetalcomments[1]{\textcolor{magenta}{#1}}

\author{\IEEEauthorblockN{Emre Karabulut}
\IEEEauthorblockA{Department of ECE\\
NC State University\\
Raleigh, NC, 27695\\
ekarabu@ncsu.edu}\vspace*{-3.0em}
\and
\IEEEauthorblockN{Arsalan Ali Malik}
\IEEEauthorblockA{Department of ECE\\
NC State University\\
Raleigh, NC, 27695\\
aamalik3@ncsu.edu}\vspace*{-3.0em}
\and
\IEEEauthorblockN{Amro Awad}
\IEEEauthorblockA{Department of ECE\\
NC State University\\
Raleigh, NC, 27695\\
ajawad@ncsu.edu}\vspace*{-3.0em}
\and
\IEEEauthorblockN{Aydin Aysu}
\IEEEauthorblockA{Department of ECE\\
NC State University\\
Raleigh, NC, 27695\\
aaysu@ncsu.edu}\vspace*{-3.0em}
        
}

\title{THEMIS: \underline{T}ime, \underline{H}eterogeneity, and \underline{E}nergy \underline{Mi}nded \underline{S}cheduling for Fair Multi-Tenant Use in FPGAs \vspace{-.5em}}

\maketitle

\begin{abstract}
Using correct design metrics and understanding the limitations of the  underlying technology is critical to developing effective scheduling algorithms. 
Unfortunately, existing scheduling techniques used \emph{incorrect} metrics and had \emph{unrealistic} assumptions for fair scheduling of multi-tenant FPGAs where each tenant is aimed to share approximately the same number of resources both spatially (in space) and temporally (in time).

This paper proposes an improved fair scheduling algorithm that fixes earlier issues with metrics and assumptions for `fair' multi-tenant FPGA use.
Specifically, we claim three improvements.
First, we consider both spatial and temporal aspects to provide spatiotemporal fairness---this improves a recent prior work that assumed all tasks would have the same latency. 
Second, we add the energy dimension to fairness: by calibrating the scheduling decision intervals and including their energy overhead, our algorithm offers trading off energy efficiency for fairness. Third, we consider previously ignored facts about FPGA multi-tenancy, such as the existence of heterogenous regions and the inflexibility of run-time merging/splitting of partially reconfigurable regions. We develop and evaluate our improved fair scheduling algorithm with these three enhancements. Inspired by the Greek goddess of law and personification of justice, we name our fair scheduling solution THEMIS: \underline{T}ime, \underline{H}eterogeneity, and \underline{E}nergy \underline{Mi}nded \underline{S}cheduling.

We implemented our solution on Xilinx Zedboard XC7Z020 with real hardware workloads and used real measurements on the FPGA to quantify the savings of our approach.
Compared to previous algorithms, our improved scheduling algorithm enhances fairness between 24.2--98.4\% and allows a trade-off between 55.3$\times$ in energy vs. 69.3$\times$ in fairness. The paper thus informs cloud providers about future scheduling optimizations for fairness with related challenges and opportunities.

%
\end{abstract}


\section{Introduction}
\label{sec:intro}
FPGAs have been recently added to the cloud, and multi-tenant use in the cloud can reduce costs, as in CPU/GPU-based cloud infrastructure.
Multi-tenancy can occur either \emph{spatially} (\textit{i.e.,} sharing different resources at the same time) or \emph{temporally} (\textit{i.e.,} sharing the same resources across various time slots) using partial reconfiguration (PR).
Multi-tenant scheduling on cloud FPGAs is especially challenging, given FPGA architecture and re-programming costs. 
Efficiency and fairness are essential goals of scheduling---while efficiency aims to use the available resources maximally, fairness aims to distribute them equally among tenants.  
This paper focuses on the fairness aspect of multi-tenant FPGA scheduling.

Unfortunately, prior works have omitted fair scheduling in multi-tenant FPGAs and instead focused on efficiency~\cite{AmorphOS,Coyote,Fahmy,fake_one,OPTIMUS,PR_OpenCL,GPU_Scheduling, GPU_Sch_survey, CPU_Sche_2, CPU_Sche,qos,PRbenefits,PR_Costs}.
A recent work, named spatiotemporal FPGA scheduling (STFS) has aimed to tackle fairness~\cite{STFS}, but it has two major issues: (i) metrics and (ii) assumptions.
Indeed, correct metrics are important to identify and measure progress toward meaningful goals, and correct assumptions about the underlying technology are important to feasibly port algorithms to real-world systems.

\emph{(i) Metric issues:} Although STFS aimed for spatiotemporal fairness, it only considers a tenant's area demand in its fairness calculation and ignores the computation time as a metric.
Therefore, a tenant with a low area but a long computation request can unfairly occupy the resources.
The energy aspect of scheduling is also ignored.
\emph{(ii) Assumption issues:}
STFS also omits the practical limitations and use-case of multi-tenant FPGAs.
First, it assumes that all shared regions are identical and can be easily merged/split at run-time, allowing for flexible re-configuration.
By contrast, commercial FPGAs have a heterogeneous distribution of BRAMs, LUTs, and DSPs in the layout and do not support dynamically adjusting PR regions (at run-time).
Second, STFS assumes that the scheduling decisions are made at fixed intervals. 
This is not a fundamental requirement and may change based on application needs.
Third, STFS assumes constant execution of all workloads, which may not always be true in real-world applications. 
Therefore, we need an improved fair scheduling algorithm that addresses these issues.

This paper proposes an improved fair scheduling algorithm that addresses issues in the earlier work.
Inspired by the Greek goddess of law and personification of justice, we name our fair scheduling solution THEMIS: \underline{T}ime, \underline{H}eterogeneity, and \underline{E}nergy \underline{Mi}nded \underline{S}cheduling.
We claim the following contributions.
\begin{itemize}

\item \textbf{Latency-aware fair scheduling.} We factor in tenants' timing/throughput requirements differences when establishing a fair scheduling policy.  We added this metric when calculating the scheduling score and optimized for it in our scheduling algorithm.

\item \textbf{Energy-aware fair scheduling.}  Rather than using fixed intervals for scheduling decisions, our algorithm can adjust it based on energy--fairness needs, allowing flexibility and trade-off in case energy is a design dimension.

\item \textbf{Heterogenous region management.} We do not assume tenant regions have the same amount of resources or that multiple regions can be flexibly merged/split at run-time.  Our proposed algorithm considers the differences in tenant regions and makes realistic assumptions when establishing a fair scheduling policy.

\item \textbf{Always vs. random demands.} We do not assume that all tenants will always demand to perform tasks all the time.  We also create a random demand test scenario with randomized tenant work requests.
We evaluate our scheduling algorithm on a Xilinx ZynQ System-on-Chip (SoC) and compare our results with prior works based on the two scenarios: tenant allocation requests appear in (i) recurring-precise order and (ii) randomized order.

\end{itemize}

Compared to previous algorithms, our scheduling improves fairness between 
24.2\%--98.4\%, and allows a trade-off between 55.3$\times$ in energy vs. 69.3$\times$ in fairness.

\begin{figure}[t]
\centering
    \hspace*{-1em}
	\includegraphics[width=0.49\textwidth]{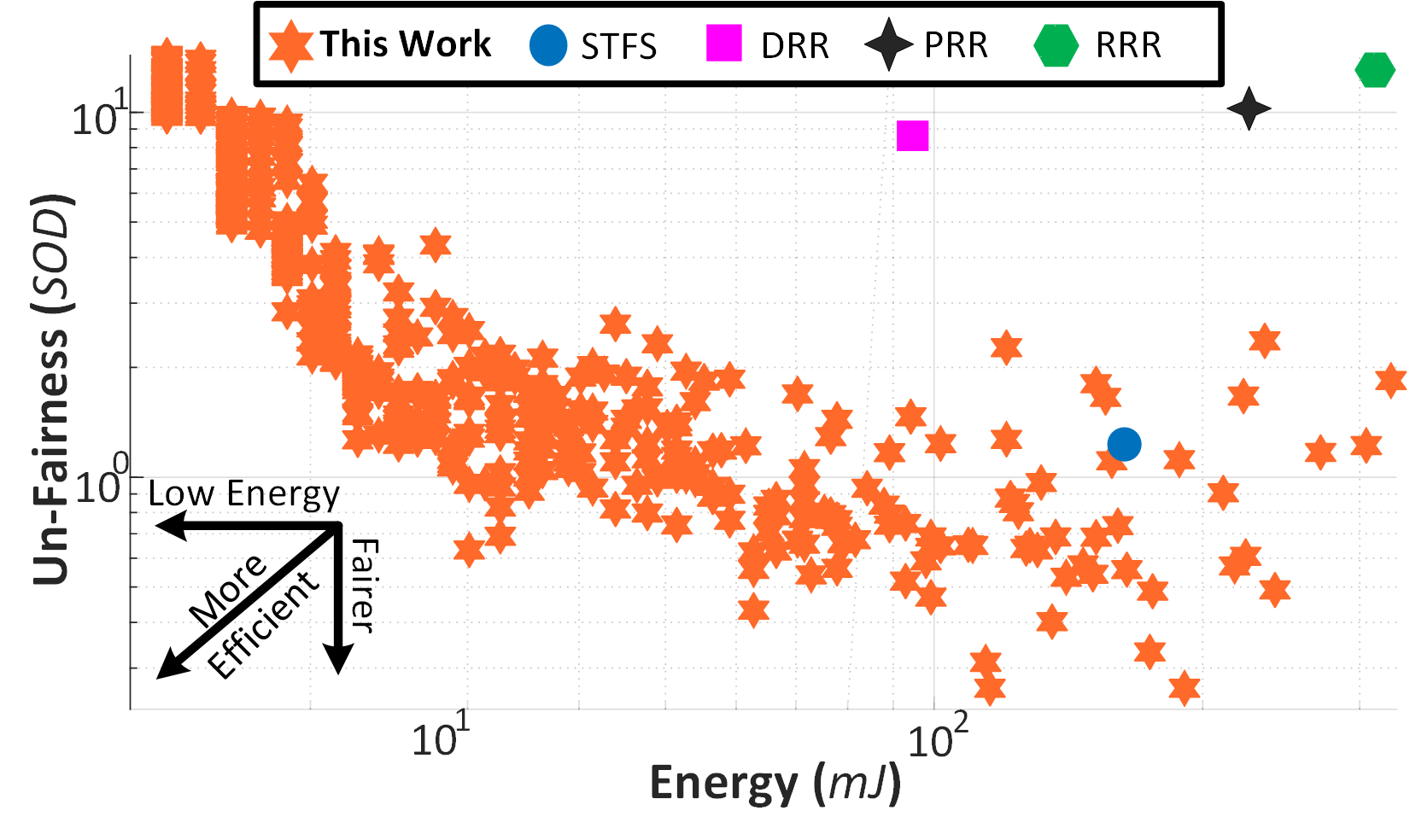}
	\vspace{-1.5em}
	\caption{Fairness and energy efficiency comparison of our work with prior for recurring-precise order (in logarithmic scale). The proposed methodology improves both fairness and energy efficiency while allowing trade-offs. 
 }
	 \vspace{-1.5em}
	\label{fig:eng-fairness}
\end{figure}

Figure \ref{fig:eng-fairness} quantifies our key claims. 
It depicts our work's fairness vs. energy efficiency and previous algorithms, including STFS~\cite{STFS}, Plain, Relaxed, and Deficit Round Robin (PRR, RRR, DRR). 
The figure shows our two advantages.
First, we offer a trade-off between energy efficiency vs. fairness by calibrating the time intervals of scheduling decisions.
Second, we improve fairness by considering the tenant's workload duration in addition to its area demand, which might result in unfair hoarding of FPGA slots if left unchecked.\\
\noindent
\textbf{Organization.} The rest of the paper is organized as follows.
Section~\ref{sec:background} outlines the existing research related to multi-tenancy for FPGAs. In this section, we identify the limitations of current methodologies and propose a new metric in Section~\ref{sec:new_metric} that overcomes the identified limitations.
Section~\ref{sec:design} presents the details of the proposed scheduling technique, THEMIS, for fair spatiotemporal tenant allocation in cloud FPGAs.
In Section~\ref{sec:results}, we provide a comprehensive evaluation of THEMIS, substantiating its effectiveness through a series of rigorous benchmarks and experimental results.
Section~\ref{sec:discussions} seeks to inspire further research in this field, articulating the need for advancements and further exploration. The paper is concluded in Section~\ref{sec:conclusions}, where we summarize our findings and propose potential directions for future work.



\section{FPGA Multi-Tenancy and Tenant Scheduling}
\label{sec:background}

The recent high-capacity FPGAs enable multi-tenancy where several tenants may share the same FPGA to increase resource utilization and reduce service costs.
Scheduling is fundamental in providing efficient, faster, and fairer multi-tenancy.
Conventional OS scheduling techniques do not easily adapt to cloud FPGA virtualization systems since FPGAs have a different architecture compared to other processing systems (\textit{e.g.,} CPUs/GPUs)~\cite{GPU_Scheduling, GPU_Sch_survey, CPU_Sche_2, CPU_Sche}.

\begin{table*}[t]
\huge
\centering
\caption{A comparative analysis of prior scheduling frameworks, their respective design objectives, and optimization approaches.
}
\label{tab: Prior_frameworks}
\resizebox{\textwidth}{!}{%
\begin{tabular}{|l|l|l|l|l|l|l|}
\hline
\textbf{Prior Work} &
  \textbf{\begin{tabular}[c]{@{}l@{}}Temporal \\ Tenancy\end{tabular}} &
  \textbf{\begin{tabular}[c]{@{}l@{}}Spatial \\ Tenancy\end{tabular}} &
  \textbf{Scheduling Policy} &
  \textbf{\begin{tabular}[c]{@{}l@{}}Demand \\ Type\end{tabular}} &
  \textbf{Optimization Goal} &
  \textbf{Scheduling Limitations} \\ \hline
OPTIMUS~\cite{OPTIMUS} &
  No &
  Yes &
  Round-robin &
  Always &
  - Minimum area utilization &
  \begin{tabular}[c]{@{}l@{}}- No spatial fairness\\ - No energy awareness\end{tabular} \\ \hline
A\textsubscript{MORPH}OS~\cite{AmorphOS} &
  Partially &
  Yes &
  Equal distribution of IO and bandwidth &
  Random &
  - Minimum area utilization &
  \begin{tabular}[c]{@{}l@{}}- No spatiotemporal fairness\\ - No energy awareness\end{tabular} \\ \hline
STFS~\cite{STFS} &
  Yes &
  Yes &
  \begin{tabular}[c]{@{}l@{}}Area aware policy for fair resource allocations \\at fixed time intervals \end{tabular} &
  Always &
  \begin{tabular}[c]{@{}l@{}}- Area-based efficient resource allocation\\ to meet desired average allocation\end{tabular} &
  \begin{tabular}[c]{@{}l@{}}- No tenant latency awareness\\ - No energy awareness\end{tabular} \\ \hline
Fahmy et al.~\cite{Fahmy} &
  Yes &
  Yes &
  \begin{tabular}[c]{@{}l@{}}No policy, serves tenants if tenants can fit into PR \\regions/slots or  until all PR regions/slots are full\\\end{tabular} &
  Not specified &
  \begin{tabular}[c]{@{}l@{}}- Virtualized communication interface for\\   integrating FPGA in the cloud using PR\\ - Increase computational efficiency over\\  software based virtualization\end{tabular} &
  \begin{tabular}[c]{@{}l@{}}- No spatiotemporal fairness\\ - No energy awareness\end{tabular} \\ \hline
Blaze~\cite{Blaze} &
  No &
  Yes &
  First-come-first-served &
  Random &
  - Reducing tenant allocation wait time &
  \begin{tabular}[c]{@{}l@{}}- No spatiotemporal fairness\\ - No energy awareness\end{tabular} \\ \hline
Vaishnav et al.~\cite{PR_OpenCL} &
  No &
  Yes &
  Jain index &
  Both &
  - Minimum area utilization &
  \begin{tabular}[c]{@{}l@{}}- No temporal fairness\\ - No energy awareness\end{tabular} \\ \hline
ViTAL~\cite{fake_one} &
  No &
  Yes &
  \begin{tabular}[c]{@{}l@{}}Communication aware policy, divides FPGA into\\ multiple identical blocks to have a homogenous \\ abstraction that can be merged on-the-fly.\end{tabular} &
  Always &
  \begin{tabular}[c]{@{}l@{}}- Efficient area utilization\\ - Fine-grained resource management\\ - Minimizing latency overhead of tenants\end{tabular} &
  \begin{tabular}[c]{@{}l@{}}- No spatiotemporal fairness\\ - No energy awareness\end{tabular} \\ \hline
  
Coyote~\cite{Coyote} &
  Yes &
  Yes &
  Round-robin &
  Random &
  \begin{tabular}[c]{@{}l@{}}- Reduced memory overhead\\ - Memory channels bandwidth optimization\\\end{tabular} &
  \begin{tabular}[c]{@{}l@{}}- No spatial fairness\\ - No energy awareness\end{tabular} \\ \hline

  \textbf{This Work}&
  \textbf{Yes} &
  \textbf{Yes} &
  \begin{tabular}[c]{@{}l@{}}\textbf{Time, heterogeneity, and energy minded} \\\textbf{spatiotemporal policy to meet average allocation} \end{tabular} &
  \textbf{Both} &
  \begin{tabular}[c]{@{}l@{}}\textbf{-  Latency-aware fair scheuling}\\\textbf{- Area-aware fair scheuling}\\\textbf{- Energy-aware fair scheuling}\\\textbf{- Heterogenous region managaemnt}
  \\\end{tabular}
  &
  \begin{tabular}[c]{@{}l@{}}\textbf{- Increased timing complexity}\\    \textbf{\hspace{.5em}compared to STFS ($\approx$10\%)}\end{tabular} \\ \hline
\end{tabular}%

}
\end{table*}

\subsection{Partial Reconfiguration and Dynamic Scalability}

Current FPGA sharing proposals~\cite{Fahmy,pr_proposals, AmorphOS} rely on PR hardware support, which enables dynamic programming of an FPGA slot (region) while the other FPGA slots remain active.
This naturally splits the multi-tenant FPGA into dynamic PR slots and a static portion.
The cloud vendor can dynamically program the PR slots for tenants while the static portion remains fixed and provides an infrastructure for reconfiguring the dynamic area.

The PR support has four practical limitations, and no prior multi-tenant FPGA scheduling work has addressed these limitations.
\\
\textbf{1) Static allocation}.\hspace{1mm} The size of a given PR slot is statically determined at design time and cannot be re-adjusted at run time.
A granularity restriction is imposed upon each FPGA family by the vendor itself.
For example, the smallest slot size should be at least one element (CLB, DSP, BRAM) wide and one clock region tall in Xilinx FPGAs~\cite{PR_book,malik2025epoch}.
\\
\textbf{2) Non-overlapped partitioning}.\hspace{1mm} No two PR slots can \textit{share} PL fabric, nor can one slot be a superset of smaller slots.
As a result, an FPGA region is assigned to one particular slot, and slot merging or expansion cannot occur dynamically.
\\
\textbf{3) Slots' resource heterogeneity}. \hspace{1mm}Two slots with the same size may have different hardware resources due to the heterogeneous distribution of FPGA resources.
Therefore, assuming that two tenants receive the same number of resources is incorrect, even if their slots' sizes are the same.
Finally, the bitstream generated for a particular slot cannot be re-used to configure any other slot, even if they have the same size.
\\
\textbf{4) Power/timing overheads}.\hspace{1mm} PR itself is a power-hungry and time-consuming process~\cite{karabulut2022pr, PR_power, malik2025epoch}.
Although context-switch (accomplished using PR) is necessary to preserve fairness in multi-tenancy, the PR operation places a considerable overhead on the system that must be accounted for.

In contrast to prior research, this work presents a fair scheduling scheme while adhering to the aforementioned practical limitations.
This effort separates our work from the prior work and results in a feasible and deployable scheduling solution for multi-tenant cloud FPGAs rather than offering a hypothetical solution.

\subsection{Fairness Drawback in Existing Multi-Tenant Scheduling}
\label{subsec: fairness}
In addition to omitting practical limitations, prior works propose either a complete lack of temporal fairness or insufficient fairness in the scheduling policies for multi-tenant cloud FPGAs.
There is a corpus of research on efficient use in multi-tenant cloud FPGAs, but they omit temporal fairness~\cite{Coyote, Fahmy,fake_one, PR_OpenCL, OPTIMUS, AmorphOS}.
By contrast, the recent work STFS~\cite{STFS} aims to provide spatial and temporal fairness by a scheduling algorithm for multi-tenant FPGAs.
However, STFS~\cite{STFS} only considers a tenant’s area demand in its fairness calculation and ignores the computation time associated with the current tenant's program.
Therefore, a tenant with a low area but a longer computation request can unfairly occupy the resources.

STFS defines the desired \emph{fair} allocation using Equation~\ref{eq: STFS} where $AA_{STFS}$ is a tenant's average allocation, $HMTA$ corresponds to ``how many times allocated" to indicate a tenant's number of task completions, while $A$ reflects a tenant's area demand and $NTI$ is the number of timing intervals~\cite{STFS}.

\vspace{-0.75em}
\begin{equation}
\label{eq: STFS}
\begin{split}
AA_{STFS} & = \frac{ (A \times HMTA)}{NTI} 
\end{split}
\end{equation}
\vspace{-0.5em}

STFS's fairness objective is to provide an equal average allocation to each tenant across the entire scheduling time.
For this purpose, STFS~\cite{STFS} presents the ``desired average allocation" term, which represents the theoretically fairest average allocation value.
Unlike average allocation, the desired average allocation is statically set before scheduling begins and does not change after each interval. 
This term determines a target for STFS's scheduling algorithm.
STFS's scheduling algorithm aims to minimize the difference between the desired and the current average allocation over the course of its process.
The desired average allocation is calculated simply by dividing the available PR area by the tenant number.
For example, if there are three tenants $T_1$--$T_3$ and a single PR slot that has $6$ area units, STFS's desired average allocation for each tenant 
would be $2$ ($\frac{6}{3}$).

We can also find how many intervals the scheduling algorithm needs in order to meet this desired average allocation per tenant.
This need can be defined by the least common multiple (LCM) value of the tenants' workloads.
Suppose that the tenants $T_1$, $T_2$ and $T_3$ have area needs of $2$, $3$, and $4$ units, respectively.
The LCM value of the tenants' workloads is $12$, which implies their desired $HMTA_i$ values are $6$, $4$, and $3$, respectively.
Therefore, STFS's scheduling algorithm needs $13$ ($6+4+3$) $NTI$ to achieve a fair distribution.

We highlight three issues that STFS neglects to emphasize the need for a better FPGA resource allocation methodology.
STFS first omits the technical limitations of the FPGA by assuming that one PR slot can be divided into multiple identical sub-slots and the tenant IPs can be merged and allocated altogether.
Second, STFS disregards the tenant's computation time when determining the desired average allocation. For example, two tenants might have the same area needs, but one has a significantly longer computation time than the other.
Third, STFS assigns the same number of sub-slots to both tenants, despite the fact that one tenant occupies the sub-slots for significantly longer than the other, resulting in an unfair allocation.
Our premise is that each tenant should pay for both their area workload and computational time load.

\subsection{Limitations of Existing Frameworks}
\label{subsec: existing_metric}

The choice of a shared multi-tenant system is often made to justify the hardware and associated power costs. The existing frameworks utilize methods such as virtualization~\cite{Fahmy, fake_one} and partial reconfiguration~\cite{PR_OpenCL} to provide a way to share such systems. These shared multi-tenant systems aim to provide a fair allocation through effective resource management but lack a tightly-knit management mechanism.  Moreover, given the heterogeneous nature of FPGA resources, the traditional management policies of comparable systems such as processors, caches, and memories do not easily adapt to FPGAs.


One proposed method for FPGA scheduling is the one-shot policy, which repeatedly makes independent allocations to achieve average distributions~\cite{CuttleSys, REF, DRF, BeyondDRF}.
A significant limitation of these policies is the omission of the tenant's computation \emph{time} as a crucial metric.
This omission results in a system where current allocations do not depend on past decisions and future allocations are not influenced by the present state.
The underlying reason for this disconnect is that the scheduling policies are chiefly guided by \textit{inflexible} optimization objectives.
Max-min fairness~\cite{CuttleSys}, resource elasticity fairness~\cite{REF}, and dominant resource fairness~\cite{DRF} serve as instances of such policies.
For a better grasp of this subject, Table~\ref{tab: Prior_frameworks} offers a succinct comparison of previous frameworks, their implemented policies, and inherent limitations.

The comparison reveals prior work's inclination towards \textit{spatial} tenancy rather than both \textit{spatial} and \textit{temporal} tenancy. Prior works such as A\textsubscript{MORPH}OS\cite{AmorphOS}, OPTIMUS~\cite{OPTIMUS}, Vasihnav, et al. \cite{PR_OpenCL} are mostly driven by the single goal of minimizing area utilization while trying to ensure fair allocation. These works' shortcomings include (i) a lack of spatiotemporal fairness in their scheduling, (ii) an inability to achieve desired allocation consistently, and (iii) excessive dependence on preemption without taking into account the associated costs such as time and energy. While more recent efforts, such as STFS\cite{STFS} and Fahmy et al.~\cite{Fahmy}, take into account both spatial and temporal aspects, they either apply no scheduling policies or apply policies that are based on the wrong metric. Additionally, these earlier studies made incorrect assumptions, such as identical slot regions and slots merging/splitting in runtime. These works also failed to account for the energy overhead associated with the PR operations and the tenant's computational time.


\section{A New Metric for Fair Multi-Tenant Scheduling}
\label{sec:new_metric}
We introduce a new metric, the tenant's computation time, to establish a fairer scheduling policy for multi-tenant cloud FPGAs. 
Unlike STFS, our policy calculates the average allocations by taking into account the tenants' computational time loads and their area demands.
For a tenant $i$, we define the workload as $A_i\times CT_i$ where $A_i$ is the tenant's area demand, and $CT_i$ is its computational time load.

Another unique attribute of our work is that our policy aims to provide flexibility in performance by offering varying interval lengths rather than one constant fixed interval length.
Therefore, we do not use $NTI$ and define a more embracing term $Total\ Execution\ Time$, which is simply $NTI \times Interval$ $Length$ in the context of STFS.

\vspace{-0.75em}
\begin{equation} 
\label{eq: our_avr}
\begin{split}
AA_{Proposed}  & = \frac{ (A \times CT \times HMTA)}{Total\  Execution \ Time} 
\end{split}
\end{equation}
\vspace{-0.5em}

Equation~\ref{eq: our_avr} shows the calculation method for our proposed average allocation. 
As the scheduler progresses, the corresponding $A_i$ and $CT_i$ remain constant for each tenant, whereas $HMTA_i$ is updated.
To find the desired $HMTA_i$, we need to calculate the LCM of all tenants' workloads, just like in the STFS calculation,
and then divide that by each tenant's corresponding workload.
However, the proposed tenant's workload definition is more comprehensive than STFS's since the proposed tenant's workload consists of both area and computational time loads.
As a result, we need to calculate the LCM of all tenants' area-time products ($A_i \times CT_i$) rather than just tenants' area demands ($A_i$).
Finally, we divide the calculated LCM value by each tenant's corresponding area-time product to obtain a tenant's desired $HMTA_i$ values.

We now consider applying this metric to assess the desired average allocation on the same example used earlier for STFS. 
Tenants $T 1$, $T 2$, and $T 3$ still have area workloads of $2$, $3$, and $4$, but they also have computational time load with $5$, $2$, and $1$.
The FPGA has one PR slot, which is a $6$ sub-slot-large but cannot be divided; hence, one slot can be used by only a single tenant.
The LCM value of the tenants' workloads is $60$, so their desired $HMTA_i$ values are $6$, $10$, and $15$, respectively.

\vspace{-1.55em}
\begin{equation} 
\label{eq:eq2}
\begin{split}
Total\  Execution \ Time  & = \sum_{i=1}^{N} {CT_i \times HMTA_i }
\end{split}
\end{equation}
\vspace{-0.95em}

Equation~\ref{eq:eq2} describes how to determine the scheduling algorithm's desired execution time to achieve the desired average allocation.
The equation first multiplies each tenant's computational time load by the required $HMTA_i$ value  and then adds the results together.
For the given example, the desired $Total\  Execution \ Time$ is $65$ ($5 \times 6+2\times10+1\times15$).
Since we know a tenant's workload, its corresponding desired $HMTA_i$ values, and the desired $Total\  Execution \ Time$, 
we can calculate the desired average allocation by using Equation~\ref{eq: our_avr}, and that gives $0.92$ ($60/65$).

\begin{figure}[t]
\centering
    \vspace{-1.5em}
	\includegraphics[width=0.48\textwidth]{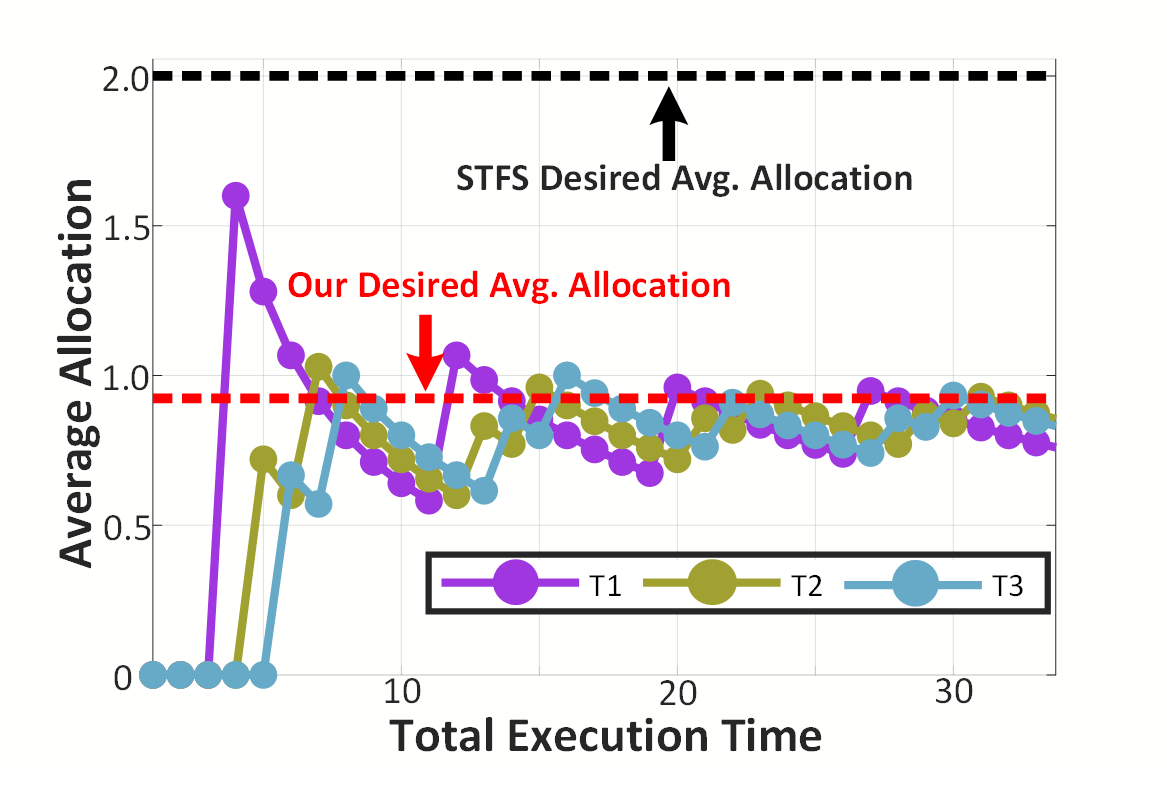}
	\vspace{-1.5em}
	\caption{The proposed work's desired average allocation (red dashed line) and STFS's desired average allocation (black dashed line). Our work takes latency into account, while STFS omits latency, and thereby its average allocation results in unfairness.}
	\vspace{-1.5em}
	\label{fig:target_alloc}
\end{figure}

Figure~\ref{fig:target_alloc} depicts the desired average allocation discrepancy between our work and STFS~\cite{STFS}.
We calculated the desired allocation for both our metric and STFS for the discussed example and then presented them in the figure as dashed red lines and black lines, respectively.
We run our scheduling algorithm and plotted a tenant's allocation over time.
The results (in Section~\ref{sec:results}) show that this change in desired average allocations corresponds to a 24\% improvement in fairness compared to STFS and up to 98\% compared to RR variants.

So far, we have discussed the case where the cloud FPGA consists of a single slot.
Therefore, the final step is to include the slot number in calculating the desired average allocation.

\vspace{-1.0em}
\begin{equation} 
\label{eq:eq3}
\begin{split}
AA_{D} & = {AA_D^{(s1)}{}_{Proposed}} \times S_N
\end{split}
\end{equation}
\vspace{-1.25em}

Equation~\ref{eq:eq3} shows the desired average allocation calculation for the general case ($S_N \geq 1$), where $S_N$ refers to the number of PR slots in the cloud FPGA.
For the general case, the desired average allocation $AA_D^{(s1)}{}_{Proposed}$ is obtained for one slot.
Then, $AA_D^{(s1)}{}_{Proposed}$ is multiplied by the number of PR slots ($S_N$).
STFS applies the same strategy for calculating the desired average allocation for multiple slots.

\vspace{.5em}
\noindent\textbf{Energy}.\hspace{1mm} In addition to the flexibility in performance, our policy provides a trade-off between energy consumption and fairness with varying interval lengths.
The interval length defines how frequently the scheduler evaluates the tenant's average allocation and performs the PR decision.
Since PR itself is a power-hungry operation, the interval length is a key factor in energy usage.
However, the prior works omit this fact.
Unlike the prior works, our work is able to take energy into account by using more embracing terms described in the previous sections and offering energy-efficient solutions.
Section~\ref{energy-vs-fairness} discusses the impact of interval length on energy and fairness in depths.








\section{The Proposed Scheduling}
\label{sec:design}

This study presents a scheduling technique for fair spatiotemporal tenant allocation in cloud FPGAs.
The proposed methodology consists of three main initiatives.
First, this technique considers both the tenants' area workload and computational time load in the scheduling policy to improve fairness.
Second, it supports varying interval lengths to reduce slots' idle times and increases resource utilization.
Third, it avoids unnecessary PR operations to reduce energy costs while maintaining fairness among tenants.


\subsection{Implementation of the Scheduling Algorithm}
Algorithm~\ref{alg:1} shows the pseudo-code of our proposed scheduling technique.
The algorithm's inputs are two data structure sets for FPGA slots ($Slots$) and tenants ($Ts$), a task queue containing tenants' request orders, and the interval length.
The algorithm has four stages: configuration, initialization, competition, and PR execution.
In the configuration stage, the algorithm initially profiles all PR slots and gathers information about the available area provided by each slot.
The subsequent step in the configuration is to profile the tenants and attain their area ($A$) and computation time ($CT$) information.
The proposed algorithm later uses these collected workloads and slots' capacity profiles to calculate the desired target allocation as defined in subsection~\ref{sec:new_metric}.
Since the configuration is a static operation, we do not present it in the pseudo-code of our proposed scheduling. 
Between steps 1 and 3 in the algorithm, our scheduling policy ensures that all tenants' average allocation is initialized with $0$.
\begin{algorithm}[t]
\small
\renewcommand{\algorithmicrequire}{{\textbf{Input}:}}
\renewcommand{\algorithmicensure}{{Output:}}
\begin{algorithmic}[1]
\Require{$Slots, \: Ts \rightarrow$ Slots and tenant data set}
\Require{$Tasks \rightarrow$  Queued tenants' orders} 
\Require{$interval\ \rightarrow$ Scheduler evaluation period}


\ForEach {$T \in  \hspace{.2em}$Ts}
    \State $T$.AvgAlloc $= 0$
\EndFor

\While{{not} $Tasks$.\texttt{Empty}}
    \State \textit{Ts}$\:\leftarrow Tasks$.\texttt{DeQueue}()
    \While{${\texttt{isEmpty}(}\textit{Slots}{)}$}
        \tikzmark{top2}
    
        \State $S   = {\texttt{SmallestSlot}}(Slots)$
        \tikzmark{right2}
                
        \State $T{.}AvgAlloc$ $  =  {\texttt{Inc}}(T.AV)$ 
{\hspace{1.5em}  Initialization}
        \State $S{.}Tenant  = {\texttt{Update}}(T)$
            
    \EndWhile
\tikzmark{bottom2}
        \ForEach {$S \in  Slots$}
    \tikzmark{top3}
        \ForEach {$T \in \mathcal Ts$}
            \If{$\texttt{Swapping}(T, S.Tenant)$}\tikzmark{right3}
                \State $T_{tmp} =  \texttt{Get}(S.Tenant)$
                \State $S.T     =\texttt{Update}(T)$
                \State $T.AvgAlloc$ $ =\texttt{Dec}(T.AV)$

                \State $T_{tmp} = \texttt{Inc}(T_{tmp}.AV)$
                    {\hspace{3em}  Competition}
                 \State $Tasks.$\texttt{EnQueue}$(T_{tmp})$
                \State \textit{Ts} $ = \texttt{Update}(T_{tmp})$
            \EndIf
        \EndFor
         
    \EndFor
    \tikzmark{bottom3}
    \ForEach {$S \in  $ \textit{Slots}}
    \vspace{-.1em}
        \tikzmark{top4}
        \If{$\texttt{isPRneeded}(S)$}
        \tikzmark{right4}

            \State ${\texttt{PerformPR}}(S)$
                    {\hspace{3.0em}  PR Execution}

        \EndIf       
    \EndFor
    \tikzmark{bottom4}
    \State ${\texttt{Wait\_Next\_Interval}}(interval)$
\EndWhile

\caption{Pseudo Code of Proposed Scheduling
}
\label{alg:1}
\end{algorithmic}
\setlength{\textfloatsep}{0.4cm}
\setlength{\floatsep}{0.4cm}
\end{algorithm}



\begin{tikzpicture}[overlay, remember picture]
\draw [decoration={brace,amplitude=0.5em},decorate,ultra thick,red]
 ($(right2)!(top2.north)!($(right2)-(0,1)$)$) --  ($(right2)!(bottom2.south)!($(right2)-(0,1)$)$);
\end{tikzpicture}

\begin{tikzpicture}[overlay, remember picture]
\draw [decoration={brace,amplitude=0.5em},decorate,ultra thick,red]
 ($(right3)!(top3.north)!($(right3)-(0,1)$)$) --  ($(right3)!(bottom3.south)!($(right3)-(0,1)$)$);
\end{tikzpicture}

\begin{tikzpicture}[overlay, remember picture]
\draw [decoration={brace,amplitude=0.5em},decorate,ultra thick,red]
 ($(right4)!(top4.north)!($(right4)-(0,1)$)$) --  ($(right4)!(bottom4.south)!($(right4)-(0,1)$)$);
\end{tikzpicture}

\vspace{-2.0em}
Our scheduling algorithm runs unless the $Task$ queue\footnote{\vspace{-0.25em}$Tasks$ queue is implemented with the last input first output (LIFO) method.}is empty, indicating that the assigned tasks are finished.
At every interval, the scheduling algorithm fetches a set of tenant requests from the $Task$ queue and then performs the initialization routine.
In the initialization stage, the algorithm places tenants in the first smallest slot available that they can fit into.
If the tenant matches a slot, its average allocation is increased based on its workload area ($A$) and time ($CT$) product.
We call this product ($A \times CT$) an adjustment value ($AV$).
Since the average allocation is the mean of tenant allocation, the $AV$'s contribution to the subsequent mean calculation reduces after each passing interval.
Therefore, the proposed scheduling algorithm revises the tenant's average allocation accordingly.
Each tenant has its own $AV$, stored in the \textit{$Ts$} data structure.
The initialization routine has two goals: (i) It assures that all PR slots are used at all times, even if there is only one tenant, eliminating the need for competing with others. (ii) It simplifies the conflict when the tenants have an equal average allocation and $AV$ in the competition stage.

After the first slot-tenant match, the algorithm enters the competition stage, where the scheduling algorithm compares the allocated and unallocated tenants.
This competition has a simple rule: a prospective tenant ($T$) acquires the slot if the previously assigned tenant ($S.Tenant$) has a higher average allocation even after $S.Tenant$'s average allocation is reduced accordingly with its $AV$.
If the tenant $S.Tenant$ loses the competition; the algorithm reduces its average allocation and then pushes back to the $Tasks$ queue.
As the prospective tenant, $T$ wins the competition against $S.Tenant$, its average allocation is increased, and its corresponding slot's record is updated with the new tenant's configurations.

The competition stage ensures that the tenant with the lowest average allocation is given the highest priority for an allocation at each interval.
If the present available capacity is insufficient to meet the needs of the first-highest priority tenant, the slot is assigned to the second-highest priority tenant.
As a result, our approach minimizes idle times in slots and improves resource utilization while maintaining fairness.

The next and final stage is to carry out the PR operation after deciding the respective slots for tenants.
The algorithm requires the PR if and only if the new tenant of a slot differs from the prior interval.
As a result, the number of PR operations is decreased, thereby improving performance and reducing energy consumption.
For example, a slot does not require a PR operation if its current tenant does not lose the competition against another tenant at the following interval.
However, STFS's algorithm makes the PR choice without considering this case.


\begin{figure}[t]
\centering
	\hspace{-.5em}
	\includegraphics[width=0.5\textwidth]{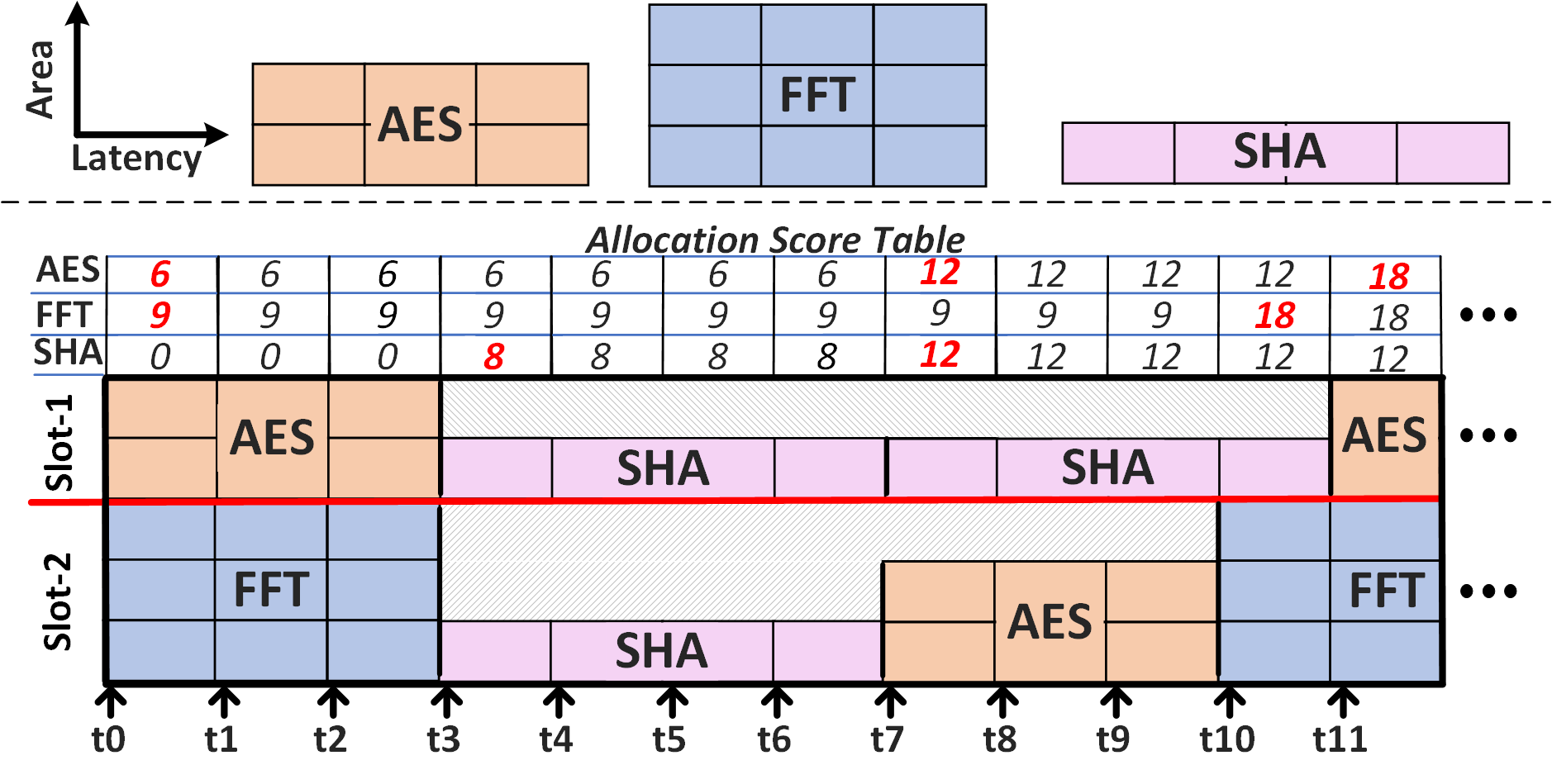}
	 \vspace{-1.0em}
	\caption{An example illustrating the slot allocation using the proposed algorithm. Three tenants: AES, FFT, and SHA, with area demand of $6$, $9$, and $4$ are competing for allocation on two FPGA slots. Slot allocation is shown for eleven intervals.}
	\vspace{-1.0em}
	\label{fig: example}
\end{figure}

\begin{figure*}[t]
\centering
\vspace{-1em}
	\includegraphics[width=\textwidth]{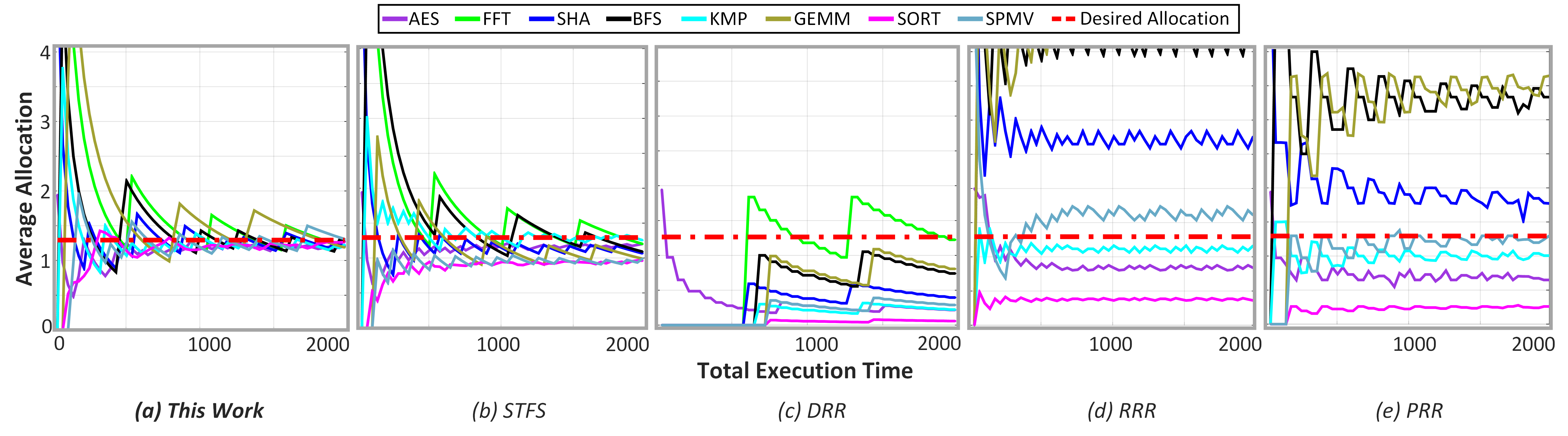}
	\vspace{-2.0em}
	\caption{Average slot allocation of proposed work on three FPGA slots with allocation interval set as $36$. (a) This Work (b) STFS (c) DRR (d) RRR (PRR). Demand for each accelerator is as per data specified in Table~\ref{tab:Resource Usage}. Desired Average Allocation is calculated as $1.243$ per tenant.}
	\label{fig:Perfromance-Evaluation}
	\vspace{-1.0em}
\end{figure*}
\subsection{Fairness and Resource Utilization}
Figure~\ref{fig: example} presents an example of our scheduling policy.
There are three tenants represented with three rectangles: AES (orange), FFT (blue), and SHA (purple).
Each tenant has a specific workload; their area workload is the height of their rectangles, while the computational time load is the width of the rectangles.
In this example, AES, FFT, and SHA require $2$, $3$, and $1$ units of area, and $3$, $3$, and $4$ units of time, respectively.
Therefore, their adjustment values ($AV$) are $6$ ($2\times3$), $9$ ($3\times3$), and $4$ ($1\times4$) units.
Additionally, the FPGA has two available slots whose area capacities are equal to the heights of the largest tenants; therefore, Slot-$1$ has $2$ area units, and Slot-$2$ has $3$ area units.
Time intervals are shown as $tn$ where '$n$' ranges from $0$ to $11$.
The tenants always have work to complete at each interval, and the scheduler carries out the tenant requests in the following order: AES, FFT, and SHA.
An allocation score table also displays tenants' allocation values as the scheduling policy progresses.
The table provides allocation values rather than averaged allocation values to avoid the complication introduced by mean computation.
Before the first allocation, the scheduler initializes all tenants' allocation numbers to $0$.


The scheduler proceeds with the given tenant order at the first time interval ($t0$); hence, AES and FFT are allocated first.
Since their $AV$s are $6$ and $9$, their allocations result in scores of $6$ and $9$.
To improve area utilization, our algorithm assigns the tenant to the smallest empty slot that the tenant can fit into; in this example, AES is allocated to Slot-$1$, whereas FFT occupies Slot-$2$.
 

Until the $4th$ interval, SHA competes with AES and FFT but cannot win the competition since the scheduler adjusts the allocation values of these two tenants with their $AV$s and obtains the same allocation value as SHA.\\
\indent
To win the competition, SHA must have a lower allocation value than the adjusted allocations of the other two tenants.
The adjusted values of these tenants (AES, FFT) are $0$, which is identical to the SHA value until the $t3$.
At the $t3$, SHA is first allocated to Slot-$1$ since it has the lowest allocation value, and its allocation score is increased by $4$.
Furthermore, because SHA's allocation score is $4$ and still lower than the other two tenants even after its first allocation, it obtains Slot-$2$ as well.
If the decision interval is larger than $3$ time-unit (in this example, it is set to $1$), AES and FFT will first start a new execution at $t3$ and then will be swapped with SHA at $t4$ without completing their work.
As a result, the execution between $t3$ and $t4$ would be wasted.
But unlike the prior work~\cite{STFS}, our algorithm can work with short intervals and minimizes resource wasting.\\
\indent
AES receives Slot-$2$ at the 8$th$ interval ($t7$)  because it has a lower allocation value than FFT and SHA.
AES is assigned to Slot-$2$ rather than Slot-$1$ because the scheduler always assigns the smaller tenant to the smaller slot.
Finally, AES first loses its slot to FFT at the $11th$ interval but then acquires Slot-$1$ by winning the competition against SHA at the $12th$ interval.


\subsection{Interval Length Effect on Energy and Fairness}
Interval is a time period where the scheduler evaluates all tenants and performs the allocation decision.
Although the prior work performs its scheduling with a fixed and long interval~\cite{STFS}, the interval length significantly affects fairness~\cite{RR_interval,1966schedule}.
Short intervals increase the scheduler's responsiveness and prompt the scheduler to meet the target fairness in a short period of time.
Yet, short intervals encourage the scheduler to perform more PR operations, consuming more energy.
Hence, the interval length is the key element in the energy and fairness trade-off. In section~\ref{energy-vs-fairness}, we discuss these elements more in-depth. 
The scheduler's ability to support dynamically varying interval lengths might come at the expense of the scheduler's performance overhead.

\section{Results and Comparison}
\label{sec:results}

\noindent
\textbf{Evaluation Platform.}  To evaluate the performance of our proposed work, we used Xilinx Zedboard XC$7$Z$020$ from the Zynq-$7000$ SoC family. We created a \textit{scheduler platform}  with the following specifications. The platform consisted of three slots, each capable of partial reconfiguration, having sizes $ S\in [4,10,18] $ on the PL side.
This is done to handle distinct, non-identical workload requirements of each tenant in Table~\ref{tab:Resource Usage}\footnote{These numbers directly correlate to the heterogeneous FPGA resources (BRAM, DSP, LUT, etc.) encompassed within a slot \textit{e.g.}, the slot size of $18$ represents having $1.8\times$ and $4.5\times$ more slice resources than that of a slot having size $10$ and $4$, respectively. 
These numbers are obtained by dividing the actual resource counts with their greatest common divisor across the slots.
}.


\vspace{-0.25em}
\begin{table}[ht]
\large
\vspace{-0.5em}
\caption{Resource and latency demands of benchmarks}
 \vspace{-0.5em}

\label{tab:Resource Usage}
\centering
\resizebox{\columnwidth}{!}{%
\begin{tabular}{|c|c|c|c|c|c|c|c|c|}\hline
  & \textbf{AES} & \textbf{FFT}& \textbf{SHA} & \textbf{BFS} & \textbf{KMP} & \textbf{GEMM} &\textbf{SORT} & \textbf{SPMV}     \\\hline
  \hline
Area & $2$& $17$& $6$& $12$&$3$& $14$& $1$& $5$ \\ \hline
Time &$7$&$5$& $8$& $15$& $9$& $28$& $14$& $14$   \\ \hline

\end{tabular}
}
 \vspace{-1.0em}
\end{table}
\vspace{.7em}
\noindent
\textbf{Workloads.} We follow the same MachSuite benchmarks in STFS\cite{benchmark}. These are canonical benchmarks used in prior works due to their diverse nature and for the emulation of real-world applications, \textit{e.g.,} AES and SHA are widely used in cryptosystems; the BFS and SORT algorithm is used to data-structure applications; GEMM and KMP utilization include machine learning and statistical analysis of complex datasets, etc.
\footnote{The choice of the benchmarks and their underlying architecture does not affect our scheduling decision. These benchmarks are merely a means to illustrate the effectiveness of THEMIS under variable workload conditions.}. Table~\ref{tab:Resource Usage} presents the area and time workloads associated with each IP within that benchmark. The original benchmarks consist of C language implementation. In order to convert them from C to RTL code, we used Vivado HLS 2020.2. After that, each IP was profiled using a two-step procedure. First, we obtained the area workload for each IP using the synthesis stage. Second, we counted each IP execution cycle by snooping the AXI interface as these IPs run on our target FPGA platform. This phase safeguards that the obtained computational time load is accurate and includes the I/O latencies for each IP.
\subsection{Performance Evaluation}
\label{Performance-Evaluation}
Figure~\ref{fig:Perfromance-Evaluation} presents our solution's fairness performance by comparing our algorithm with existing algorithms, specifically STFS and the three Round-Robin (RR) variants. For the detailed description of RR variants, we refer interested readers to STFS~\cite{STFS}.
The X-axis captures the $Total\  Execution \ Time$  of each algorithm, whereas the Y-axis represents the average slot allocations per interval.

The desired average allocation (shown in red dashed line) is $1.243$, computed using Equation~\ref{eq: our_avr} and~\ref{eq:eq2}.
We chose the scheduling interval length as $36$-time units.
Although our scheduling algorithm delivers better fairness and performance with a shorter interval length, the compared algorithms cannot schedule tenants with a shorter interval than $36$-time units.
This interval selection ensures that the other algorithms have a chance to carry out their workload execution.\\
\indent
The results demonstrate that all three RR variants sporadically deviate from the desired allocation during each interval. In comparison, STFS met the target for some tenants but missed for others (\textit{e.g.,} SHA and SPMV).
STFS performs fairly well in contrast to RR variants because it iteratively tries to minimize the gap between the current allocation and the desired allocation, but it still lacks in terms of fairness because it does not use the right metrics.
\\
\indent
STFS strays from the long-term allocations goals (see spikes at intervals of $700$ and $1300$).
This is due to the fact that STFS only considers the area demands of workloads when scheduling, neglecting the workload's computation time.
The proposed work schedules workloads by taking into consideration both area and computation time. Based on these metrics, the proposed work computes the average allocation in each interval to minimize the divergence between average and desired allocation.

The RR variants veered from the desired allocation (see the red dashed line in Figure~\ref{fig:Perfromance-Evaluation}). STFS performed better than RR variants; however, almost half of the workloads failed to reach the desired allocation. (\textit{e.g.,} SORT, AES, SHA, and SPMV fail to reach the desired allocation throughout their execution).
Even though the other four algorithms were favored by the time interval we set, our solution still outperformed them in terms of fairness by $82.0\%-98.4\%$.

\subsection{Energy and Resource Utilization Performance}
\label{energy-vs-fairness}
Our scheduling algorithm can tune the scheduling based on energy--fairness needs of the system.
Figure~\ref{fig:eng-fairness} presents our policy's performance for energy and fairness for different interval lengths and compares this performance with other scheduling algorithms.

The Y-axis of the figure represents the unfairness, which is quantified with the absolute sum of the difference (SOD) between the tenant's average allocation and the desired allocation in log scale.
The X-axis depicts the energy consumption (also in log scale), which is computed by multiplying the PR's energy need with the number of PR completions in the overall scheduler progression time.\\
\indent
Short intervals provide a fairer result for our algorithm, while long intervals optimize energy efficiency. Initially, we varied the interval range between 1--72 to observe the scheduling policies' behavior. However, eventually, we fixed the interval value at 36 to provide the results of Figure~\ref{fig:Perfromance-Evaluation}.  This is due to the lack of variable interval support in prior work. Setting the interval value lower than 36 prevented previous work from executing some of the workloads. 

The absence of variable interval support is primarily why they provide fixed energy and fairness results. 
For the fairness of 160 \textit{SOD}, our work consumes 0.33\textit{mJ} of energy, whereas STFS consumes 1.20\textit{mJ} energy. Conversely, keeping the energy constant at 1.10\textit{mJ}, the proposed work's SOD is 161, and STFS's SOD is 191. Note that higher SOD values indicate the scheduling policy's elevated unfairness.
\begin{figure}[t]
\centering
	\includegraphics[width=0.45\textwidth]{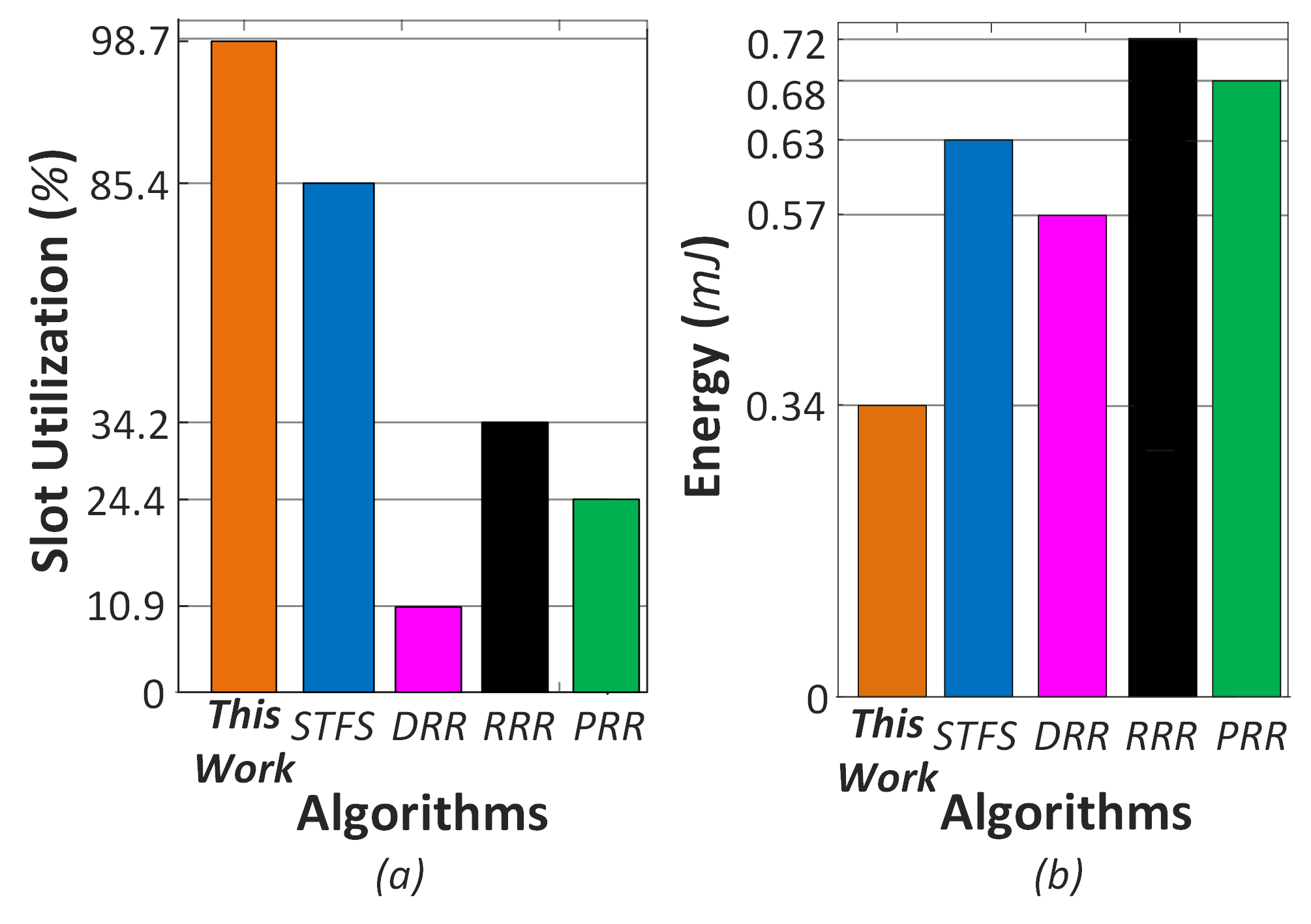}
 \label{fig: SlotUtilization}
        \caption{The scheduling algorithms' results for (a) FPGA slot utilization and (b) energy cost. Our solution reduces the slot idle time by supporting short interval lengths. The energy savings are achieved by lowering needless PR operations while scheduling. }
\end{figure} 
\\
\indent
Our solution improves slot utilization and lower energy usage compared to prior work. Calibrating scheduling intervals leads to the minimization of PR operation, which directly lowers the energy consumption of the proposed work.
Figure 5 presents a performance comparison between our and other algorithms in terms of slot utilization and energy utilization.
Figure 5 (a) shows that our solution reduces the slot idle time from $89.1\%$ to $1.3\%$on average.
Moreover, this work is not only resource-efficient but also more energy efficient.

Figure 5 (b) portrays that our solution saves energy up to $52.7\%$ due to our energy-aware scheduling policy. 
This saving is a direct result of minimizing unnecessary PR operations in scheduling, which prior efforts did not take into account.
We referred to the energy data associated with PR operations presented in~\cite{PR_power}. The bitstream size (in KB) of our PR slots are $1180$, $1340$, and $837$, each consuming 1.25\textit{mJ} on average. The total occurrences of PR operations for each algorithm were acquired from the experimental setup itself.

\begin{figure}[t]
\centering
    \includegraphics[width=0.45\textwidth]{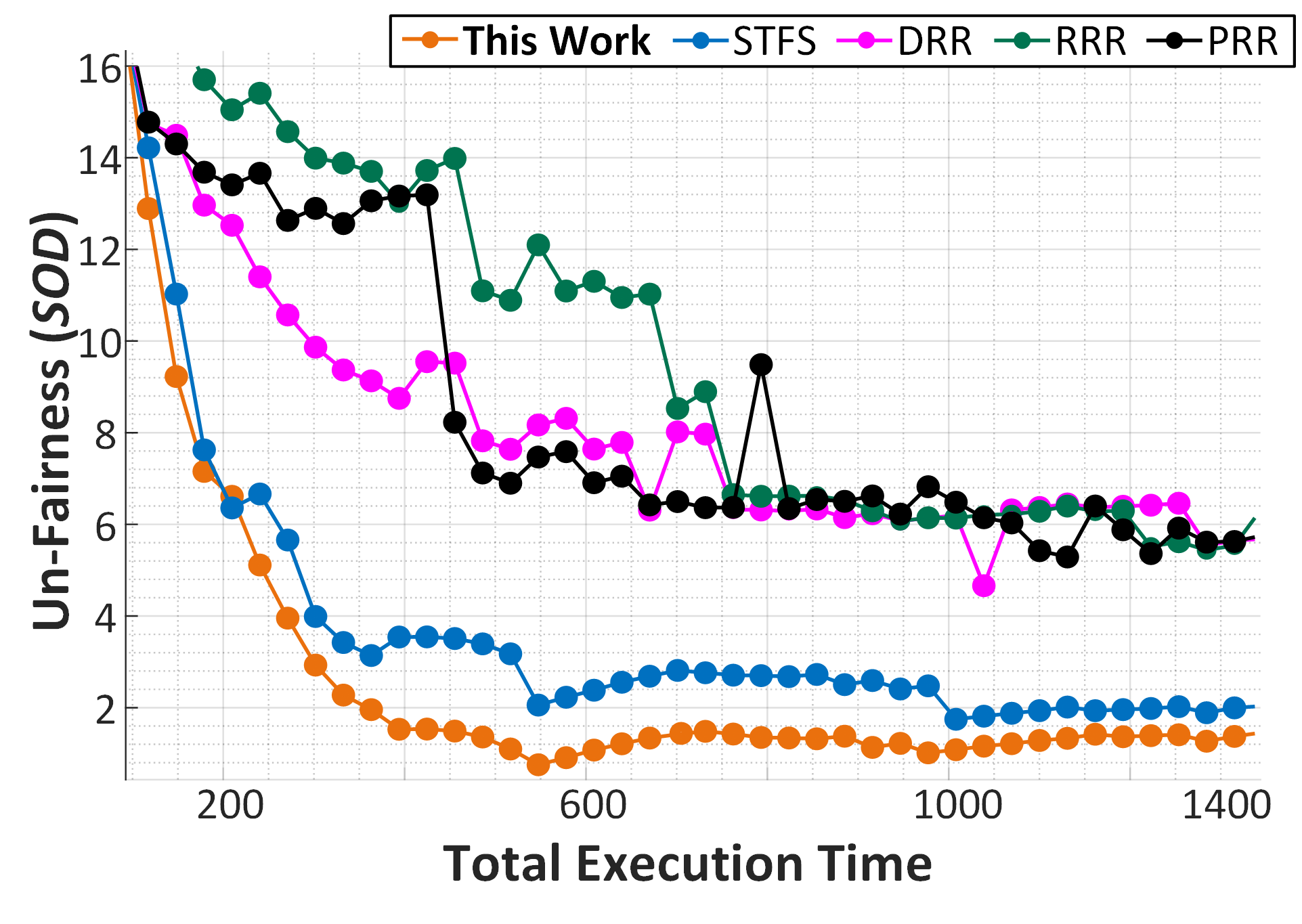}
	\caption{Unfairness in the allocation of slots is shown for the proposed and prior work in an always-demand workload. Prior work fails to meet the desired allocation, resulting in elevated levels of unfairness (measured in SOD). }
	\label{fig:SOD-with-Always-Order}
\end{figure}
\subsection{Always vs. Random Demands}
\label{AlwaysvsRandom}
The assessment reported thus far reflects the outcomes of tests based on the always-demand workload. In this scenario, the tenants always have work to complete at each interval, and their demands are submitted to the scheduler in the same order. In reality, multi-tenant cloud FPGAs may host a variety of users, resulting in various scenarios.
Therefore, scheduling algorithms should provide versatile frameworks that can still deliver efficiency and fairness under varying conditions.

The prior work ignores the case where the tenants can demand the slots in random order instead of the recurring precise-order scenario (discussed so far).
Also, a tenant can demand two slots at one interval while it can skip requesting a slot altogether in successive intervals.
Therefore, we tested and evaluated our algorithm's performance under two different scenarios: always demands and random demands.

\noindent\textbf{Always-demand}. This evaluation is beneficial because slots are finite; providing a slot to all tenants at each interval is impossible. 
Figure 6 provides a detailed comparison of the fairness of our work and previous efforts on the always-demanded workload.
The X-axis displays $Total\  Execution \ Time$, while the Y-axis depicts the unfairness in terms of SOD. 
The higher the value of SOD, the worse the algorithm performs in terms of fairness.

In this test case, PRR, RRR, and DRR failed to minimize the aggregate unfairness throughout their executions. 
STFS performs well in comparison to the three RR variants and almost comes closer to our work (see the execution time between 1000 -- 1400). 
This close performance is, however, expected because this work predicates the concept given in STFS. 
Despite this close performance, our suggested approach significantly outperforms all prior efforts, proving it to be a fair and effective solution.
\noindent\textbf{Random-demand}. In a multi-tenant cloud environment, it is pertinent to consider and adjust to the scenarios where one tenant does not currently have any tasks to execute.
A tenant may even demand more slots than others in successive intervals; thus, the order of execution can change more rapidly.
\begin{figure}[t]
\centering
	\includegraphics[width=0.45\textwidth]{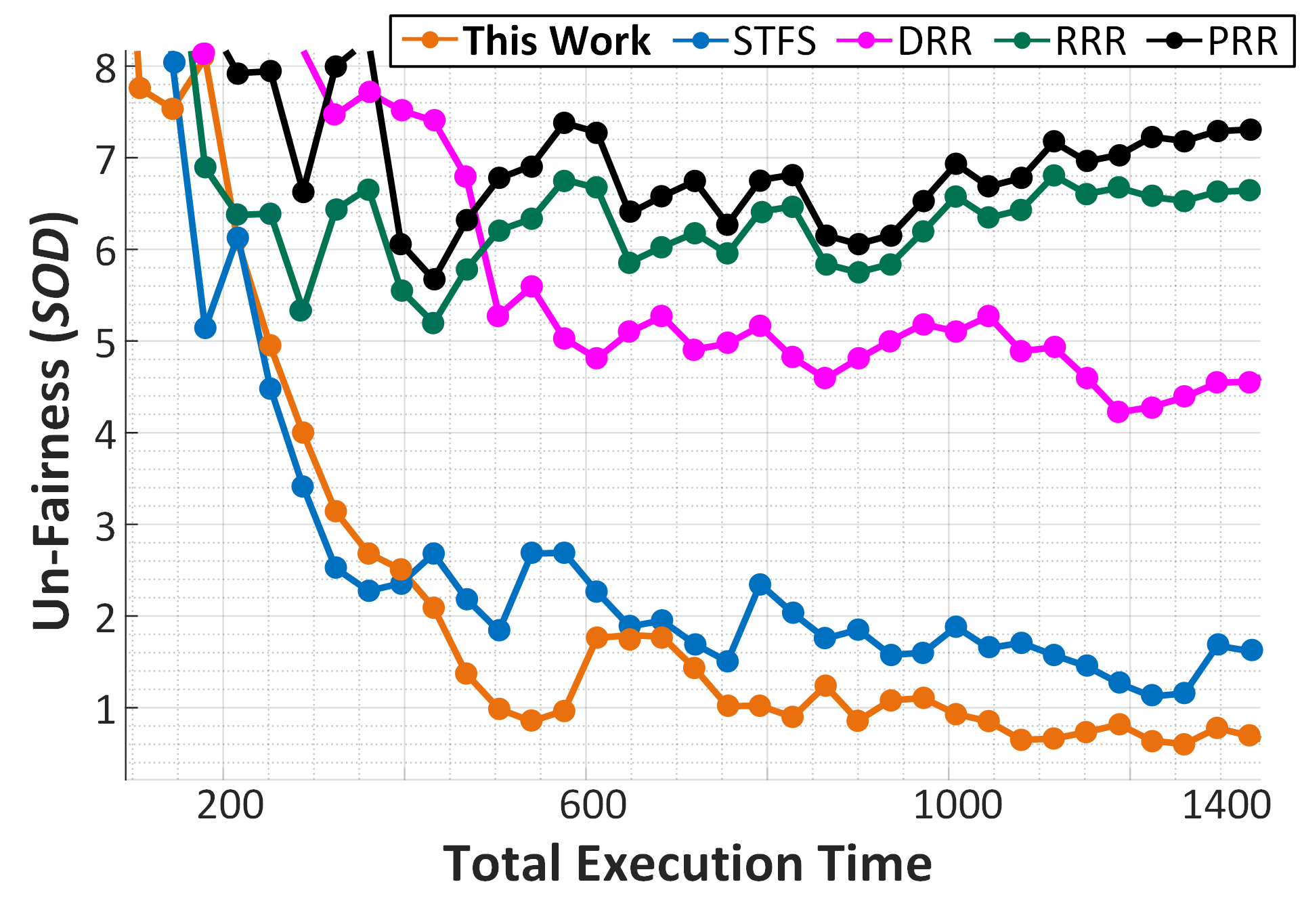}
	\caption{A comparative analysis of our proposed algorithm with prior work, utilizing random-demand tests to evaluate its fairness. The proposed algorithm demonstrates rapid recovery times without sacrificing fairness (lower unfairness represented as SOD). The tests were performed using short intervals to mimic fast incoming requests.} 
	\vspace{-1em}
	\label{fig:SOD with Random IP Order}
\end{figure}
A fair scheduling policy must handle and adapt quickly to such requests. The proposed work supports short intervals allowing it to re-evaluate and recover from deviations from the desired allocation, resulting in a response to such requests. 
Figure~\ref{fig:SOD with Random IP Order} presents the fairness performance of the proposed work and the prior works for short intervals and random-demand workloads.
The X-axis shows the $Total\  Execution \ Time$, and the Y-axis provides the un-fairness expressed as the difference in the average sum of allocations per interval, for each algorithm.\\
\indent
The definition of PRR, RRR, and DRR specified in prior work~\cite{STFS} dictates that these algorithms try to stick closely to an  ordered fashion as a result, in this case of random-order requests, their scheduling fairness is excessively distorted.
This effect is further amplified in the subsequent intervals. 
STFS, while trying to converge around the desired allocation, also deviated from the target too often. 
STFS works using large fixed intervals, which is evident from its slow response time in random demand test cases. Any random requests have a significant impact on the STFS's ability to reach the desired allocation.   
Our proposed algorithm makes adjustments to stick as closely to the desired allocation as possible while maintaining its fairness. The short intervals help the proposed work to recover quickly while maintaining fairness. 
For random-order demands, our algorithm improves the fairness between $24.2-93.1\%$ compared to prior work.

\begin{table}[b]
\large
\centering
\vspace{-1.25em}
\caption{Time to Completion for THEMIS and STFS Analyzed under variable Workload and Demand Conditions}
\label{tab: Execution_Time}
\resizebox{\columnwidth}{!}{%
\begin{tabular}{|c|c|c|c|c|c|}
\hline
\multicolumn{1}{|l|}{\textbf{\begin{tabular}[c]{@{}l@{}}Case\\ Study\end{tabular}}} &
  \multicolumn{1}{l|}{\textbf{Algorithm}} &
  \textbf{\begin{tabular}[c]{@{}c@{}}Workload\\ Area Demand\end{tabular}} &
  \textbf{\begin{tabular}[c]{@{}c@{}}Demand \\ Type\end{tabular}} &
  \multicolumn{1}{l|}{\textbf{\begin{tabular}[c]{@{}l@{}}Clock\\ Cycles\end{tabular}}} &
  \multicolumn{1}{l|}{\textbf{\begin{tabular}[c]{@{}l@{}}Time\\ (ms)\end{tabular}}} \\ \hline
\multirow{4}{*}{1} & THEMIS & \multirow{4}{*}{$ S\in [4,10,18] $} & \multirow{2}{*}{Always} & 665,349 & 2.00  \\ \cline{2-2} \cline{5-6} 
                   & STFS   &                          &                         & 606,635 & 1.82 \\ \cline{2-2} \cline{4-6} 
                   & THEMIS &                          & \multirow{2}{*}{Random} & 976,030 & 2.93 \\ \cline{2-2} \cline{5-6} 
                   & STFS   &                          &                         & 910,650 & 2.73 \\ \hline
\multirow{4}{*}{2} & THEMIS & \multirow{4}{*}{$ S\in [17,17] $}   & \multirow{2}{*}{Always} & 668,348 & 2.01 \\ \cline{2-2} \cline{5-6} 
                   & STFS   &                          &                         & 607,275 & 1.82 \\ \cline{2-2} \cline{4-6} 
                   & THEMIS &                          & \multirow{2}{*}{Random} & 960,151 & 2.88 \\ \cline{2-2} \cline{5-6} 
                   & STFS   &                          &                         & 908,742 & 2.73 \\ \hline
\end{tabular}%
}
\end{table}
\subsection{Energy vs Fairness Tradeoff}
Figure~\ref{fig:eng-fairness} illustrates the trade-offs introduced by THEMIS’s interval-based scheduling mechanism, with a specific focus on energy consumption and fairness. These metrics are evaluated under two boundary conditions: (i) the configuration using the shortest scheduling interval, which maximizes fairness but incurs the highest energy cost due to frequent partial reconfiguration (PR), and (ii) the configuration using the longest interval, which minimizes energy consumption at the expense of fairness.

The energy factor is computed as the ratio of energy consumed in the long-interval scenario to that in the short-interval scenario. The experimental measurements produced energy values of 625.2480\textit{mJ} for the longest interval and 11.2997\textit{mJ} for the shortest interval. The resulting ratio, approximately 55.3$\times$ ($625.2480/11.2997$), reflects the energy trade-off factor in short vs. long intervals. These energy measurements are illustrated---as orange hexagrams---in Figure~\ref{fig:eng-fairness} using a logarithmic scale. 

By contrast, the fairness factor quantifies the difference in resource allocation fairness by computing the ratio of the SOD between the longest and shortest scheduling intervals. The highest observed SOD ($15.0545$) occurred under the longest interval, while the minimum ($0.2170$) was achieved with the shortest interval, yielding a 69.3$\times$ fairness trade-off ($15.0545/0.2170$), as shown in Figure~\ref{fig:SOD-with-Always-Order}. Together, the 55.3$\times$  and 69.3$\times$  factors empirically substantiate the dual impact of scheduling interval selection, illustrating the scheduler’s tunability and the resulting trade-offs between energy efficiency and fairness\footnote{The energy and fairness trade-off measurements correspond to the execution time baseline reported in Figure~\ref{fig:eng-fairness} and Figure~\ref{fig:SOD-with-Always-Order}, respectively.}.

\vspace{-.25em}
\subsection{Homogenous Slot Evaluation}
\label{Homogenous-Slot-eval}
In order to assess the effectiveness of our work under rigorous conditions, we conducted an experiment with equal slot sizes. This experiment was conducted to observe the behavior of the scheduling policies when the slots may exhibit homogenous behavior. To ensure consistency, we kept the benchmarks the same as before.
We reduced the number of FPGA slots from three to two, now having the size of $ S\in [17,17] $ on the PL side. The slot size of $17$ was set to meet the area requirements of the largest benchmark, FFT.

\begin{figure}[t]
\centering
	\includegraphics[width=0.45\textwidth]{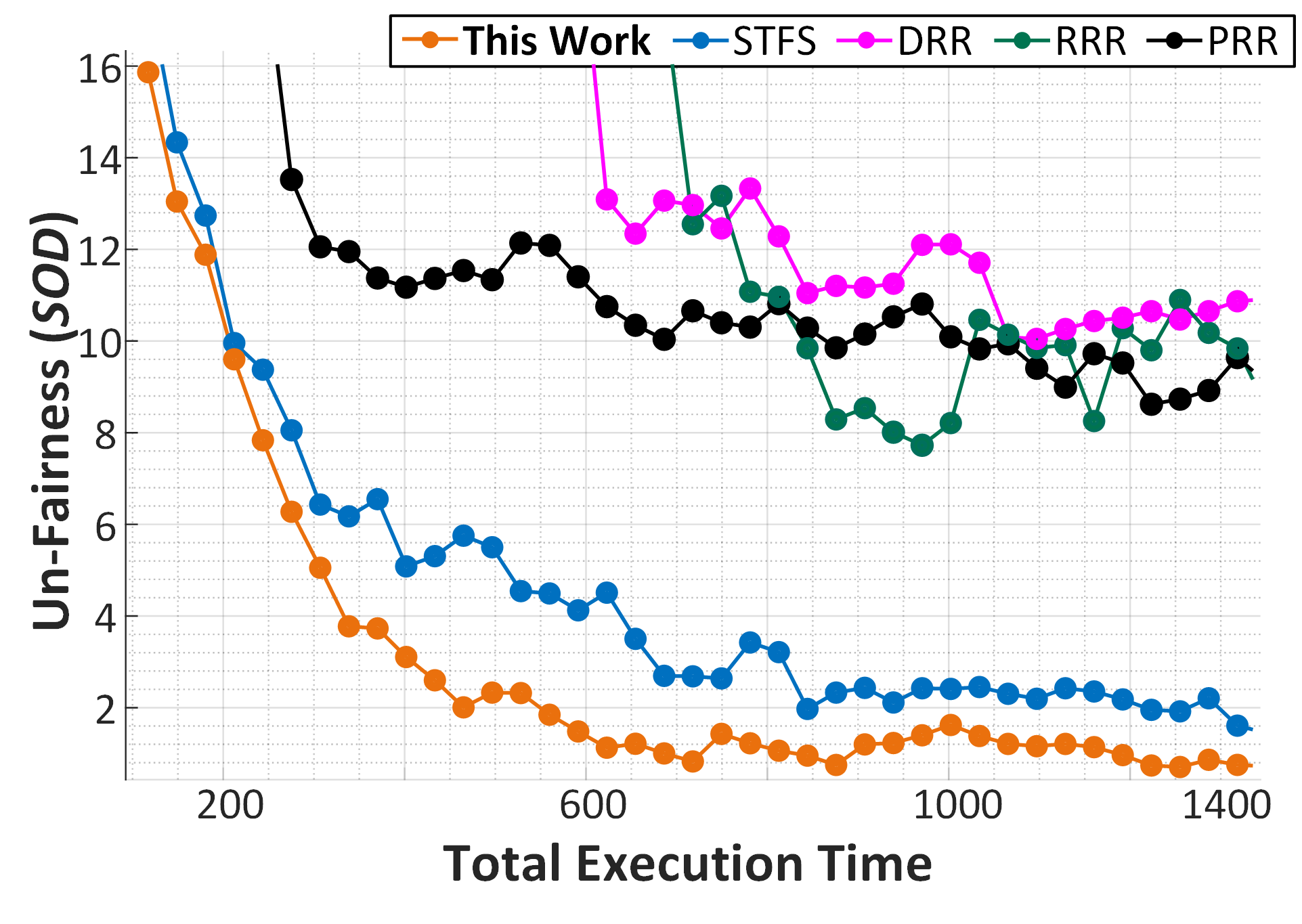}
	\caption{The proposed work was able to maintain fairness even in a scenario with limited resources, as seen in the second case study, where only two slots were available. The proposed work efficiently allocated resources and ensured fairness among all workloads.}
	\label{fig: SOD with 2 slots and random order}
\end{figure}

We conducted the experiment using the random-demand test case.
Figure~\ref{fig: SOD with 2 slots and random order} illustrates the result of this test case, with the XY-axis kept the same as the figure in section~\ref{AlwaysvsRandom}.
We observed that PRR, RRR, and DRR performed significantly worse under these restrictive conditions in terms of fairness. Furthermore, the reduced number of slots hindered in fair allocation of slots for these algorithms (also evidenced by their elevated SOD levels of 10.2, 7.8, and 11.2, respectively). While STFS closely followed the proposed work, it also failed to reduce the difference in unfairness beyond the SOD of 2.1.
This is due to the fact that STFS neither considers computation time in its scheduling policy nor supports short intervals. In contrast, the proposed work successfully lowered the SOD to 1.1, exhibiting a lower unfairness.

\subsection{Timing Overhead Analysis}\label{Timing_Analysis}
Table~\ref{tab: Execution_Time} lists the computational statistics of the proposed work for optimally distributing PL resources to each tenant for the two area workloads described in Sections~\ref{Performance-Evaluation} and~\ref{Homogenous-Slot-eval}. The comparison includes (i) always and random demand requests, (ii) execution time, and (iii) the number of clock cycles required to complete both test cases. The scheduling algorithms in both test cases are executed on the PS side at a default clock frequency of 667 MHz. A brief comparison reveals that THEMIS has approximately $10\%$ longer time to completion than STFS. However, THEMIS compensates for this overhead by offering more \textit{fairness} in its allocations compared to prior work, specifically 24.2 -- 98.4\%. Furthermore, THEMIS offers calibration of energy consumption and fairness considerations in its computations, which prior works failed to consider. This secondary feature makes THEMIS an energy-efficient and power-saving solution. In order to enhance the timing efficiency, potential usage of THEMIS includes deployment using a multi-threaded approach that leverages the dual cores of ARM Cortex-A9 on the PS side to handle tenant allocations, effectively reducing the time required for completion by a factor of two.   

\section{Discussions}
\label{sec:discussions}
This section discusses the scope of the proposed work. Furthermore, we look at some of the challenges and practical limitations of this effort.  

\subsection{Proposed Work Genericness}

In our experiments, we tested and evaluated the fairness performance of our method with two slot-size configurations.
The first configuration arranges slots in three different sizes, while the second configuration provides two slots of the same size. 
The results show that our method is extendable to different configurations having diverse slot areas.
Although fine-tuned slot-size choices may improve fairness, optimal design space exploration is out of scope.
\subsection{Partial Reconfiguration; Cost and Considerations}
The proposed work uses a PR framework to provide a state-of-the-art scheduling policy.
This framework is technology dependent and comes with different costs and latencies based on the FPGA technology.
Likewise, in practical applications, tenants may exhibit different latencies.
THEMIS provides a migratable and generic solution for spatiotemporal scheduling.
Therefore, we use the same tenant benchmarks and FPGA as STFS~\cite{STFS} to give first-order comparisons between our work and existing works.
Our contribution does not aim to improve the PR framework itself.
There is already a significant research effort to improve the performance of PR and reduce its latency~\cite{zycap,DMAICAP,vr-zycap,liu2009run,runtimePR,vipin2012high,duhem2011farm,liu1,bucknall2023zypr,malik2024enabling}.
Rather, THEMIS's scheduling policy is an orthogonal initiative that could be integrated into existing efforts. 

\subsection{THEMIS's Scalability}


The implementation of THEMIS operates independently of FPGA's logic resource constraints, including elements like LUTs, BRAMs, and FFs, among others. We exemplify the performance of our scheduling policy on a Xilinx SoC FPGA, where the scheduling algorithm operates on the PS side, and the tenants reside on the PL side. Consequently, the implementation of THEMIS eliminates the need for PL resources.
Hence, the design philosophy proposed is entirely \emph{scalable} and ideally applicable to any FPGA.
Additionally, THEMIS exhibits the ability to balance energy-vs-fairness.
While not bound by PL resource limitations, it's noteworthy that THEMIS does factor in constraints such as FPGA size, LUT quantity, and the number of tenants for \textit{fair} tenant schedules.

We used the Xilinx Zynq-7000 SoC (XC$7$Z$020$) due to its broad adoption in academic research on multi-tenant scheduling, its on-chip processor for runtime control, and native support for partial reconfiguration. Several works in reconfigurable computing have used this SoC to study scheduling, partial reconfiguration, and multi-tenancy, as it represents “mid-range” FPGAs, which, despite smaller resource counts, support the same partial reconfiguration technology as larger FPGAs~\cite{vr-zycap,zycap,PR_power,malik2020isolation,malik2025epoch}. Thus, using the Zynq SoC for evaluation follows an established paradigm.

\subsection{Time Complexity}
The determination of the time complexity of the THEMIS algorithm can be derived through an examination of its computational steps. The initial iteration (lines 6 -- 10 in algorithm 1), which continues as long as there are empty slots, entails updating tenants for said slots. The following loop sequentially iterates through each slot $S$ in the set of slots (lines 19 -- 32). For each slot, it executes another loop that iterates through each tenant $PT$ in the set of tenants (lines 21 -- 31). The complexity of this loop can be represented as $\mathcal{O}(n*m)$, where $n$ denotes the number of slots and $m$ represents the number of tenants due to its hierarchical structure. The concluding iteration traverses through each slot $S$. Assuming that the number of tenants and the slot size can be considered equivalent or scaled together within the given context, the dominant factor in our complexity analysis is the nested loop, which results in a complexity of $\mathcal{O}(n*m)$. Consequently, we can approximate that the overall complexity of THEMIS remains the same as the prior work, STFS~\cite{STFS}, \textit{i.e.,} $\mathcal{O}(n^2)$.

The current implementation of THEMIS executes and iterates through the loop serially, laying the foundation for a robust scheduling policy that balances fairness and slot utilization. While the serial lookup of slots is already effective, there is potential for further acceleration through advanced techniques such as loop unrolling and loop fusion, etc. These optimizations, although beyond the scope of the current implementation, represent exciting opportunities for future enhancements of THEMIS.

\subsection{Real-World Adaptability}
THEMIS is designed to work well in real-world systems where multiple users share the same FPGA, especially in cloud platforms like Amazon Web Services or Microsoft Azure Cloud~\cite{AWS,Azure,CloudReview}. These platforms let users run different tasks on the same FPGA, which creates challenges such as, how to divide resources fairly, reduce delays when switching tasks, and handle jobs of different sizes and types. THEMIS helps solve these problems through a smart scheduling algorithm that can adjust how and when tasks are assigned, ensuring every user gets fair access.

To emulate cloud-like conditions, THEMIS was deployed on a mid-range FPGA configured with heterogeneous reconfigurable regions/slots, capturing resource fragmentation (commonly) observed in multi-tenant use cases~\cite{dessouky2021sok,malik2024enabling}. The design supports dynamic task allocation and runtime adaptability, making it suitable for scenarios such as ML inference-as-a-service, streaming analytics pipelines, FPGA-accelerated microservices, and virtual FPGA (vFPGA) management. THEMIS enables efficient utilization of physical resources while preventing starvation and managing energy consumption, even under fluctuating load conditions. While we tested THEMIS on a smaller FPGA chip, THEMIS can fundamentally work with any size or type of FPGA. This flexibility makes it an ideal fit for future cloud systems, where it can play a key role in managing multi-user access to FPGA resources fairly and efficiently.

\section{Conclusions}
\vspace{-.25em}
\label{sec:conclusions}
This paper demonstrates that fair scheduling in multi-tenant cloud FPGA is a multi-objective, complex problem that requires considering practical issues of non-identical tenant regions, interval decision time, energy efficiency, tenant workload requirements, and random execution requests into account.  We argue that the average allocation definition should incorporate such constraints, and we propose a new algorithm that can optimize fairness under this new metric.  The proposed solution has been prototyped on a recent Xilinx FPGA SoC and quantified to improve the results achieved in prior works.  The paper, therefore, informs system designers and cloud providers about future scheduling optimization to offer fairness along with the related challenges and opportunities.
 \vspace{.5em}
\noindent \textbf{Code.} \hspace{.5em}  The code for THEMIS is open-sourced and available on \href{https://github.com/aamalik3/THEMIS.git}{GitHub} under the MIT License\footnote{https://github.com/aamalik3/THEMIS.git}.
\section{Acknowledgements}
This work was funded through the Office of Naval Research (ONR) grant N00014-21-1-2809. The views, opinions and/or findings expressed are those of the authors and should not be interpreted as representing the official views or policies of the Department of Defense or the U.S. Government.\\




\bibliographystyle{IEEEtran}
\bibliography{References}

\end{document}